# Nano-Antenna Directivity for Electromagnetic Propagation in WBANs

**Sarah Hussein[1], Walaa Sahyoun[1], Youmni Ziade[1], Raed M. Shubair [2]**

[1]Electrical and Computer Engineering Department, Beirut Arab University, Lebanon

[2]Research Laboratory of Electronics, Massachusetts Institute of Technology, USA

**Abstract** — In-vivo sensing, diagnosis and treatment of diseases is having a great attention lately. With advanced computational systems, the processing of the biological data as well as the prediction of diagnosis is becoming more promising. However, the implementation of these systems inside the human body has a major challenge: modeling the communication channel. To overcome this problem, researchers are investigating the main factors that define the characteristics of the communication channel between nano-devices. In this work, we summarize the elements that contribute to the path loss encountered by an EM wave traveling in water, skin or epidermis. Then, the impact of nano-antenna directivity on the EM propagating wave is studied along with the frequency and the communication distance. The simulation results show that the nano-antenna directivity seems to have minor contributions (5-7 dB) on the total path loss inside the human body with respect to the distance (20–30 dB) and frequency (10-15 dB).

## I. INTRODUCTION

Today, in-vivo sensing is becoming the de-facto of the biomedical engineering applications. Nano-sensors and nano-devices are the key factors of this evolution. Using these technologies, various tools are developed to control multiple entities at the atomic as well as the molecular scales [1]. Fast and accurate disease diagnosis as well as treatment is now possible thanks to the real time operational capability of nano-sensors inside the human body [2].

Despite the fact that the research on nano-devices is having good advancements, the communication between nano-sensors within an in-vivo Wireless Body Area Network (iWBAN) remains a big challenge. Mainly, three models are utilized to undergo the communication process inside the human body: ultrasound (US) communication [3], molecular communication [4] and electromagnetic (EM) communication [5]. In a fashion similar to [16-29] and [30-48], the work in this project is a continuation of the several previous contributions which have been reported in the literature in various relevant technologies, systems, and their associated applications.

The scope of this paper is to evaluate the losses encountered in the EM communication channel operating in-vivo at THz frequencies by means of distance, frequency and nano-antenna directivity. This evaluation is based on the state-of-the art models discussed in literature. In Section II, we present the path loss elements encountered in THz band and its relation to frequency, distance and nano-antenna directivity. In Section III, we present the simulation results for communication losses in water, skin and epidermis. In Section IV, we interpret the results. Finally, we draw our conclusions in Section V.

## II. PROPAGATION LOSSES

The main factors that affect an EM wave that travels at the THz frequencies are: the spreading of the propagating wave ($PL_{Spr}$), the molecular absorption from the human tissues ($PL_{Abs}$), and the scattering from the biological particles ($PL_{Sca}$). Therefore, the total path loss ($PL_t$) is given by

$$PL_t(dB) = PL_{Spr} + PL_{Abs} + PL_{Sca} \quad (1)$$

### A. *Spreading Loss*

Spreading loss occurs as the EM wave travels through the medium for a certain distance; it still diminishes as it spreads out. Assuming a spherical propagation and taking into account the nano-antenna gain effect, the spreading of the EM wave is given by the inverse-squared distance function as

$$PL_{Spr} = -10 log_{10}\left[D\left(\frac{\lambda_g}{4\pi d}\right)^2\right] \quad (2)$$

where d is the path length, $\lambda_g$ is the medium wavelength and D is the nano-antenna directivity. Basically, $\lambda_g$ is the ratio of the free space wavelength $\lambda_0$ to the real part of the refractive index of the material $n'$, such that $\lambda_g = \lambda_0/n'$.

From [6], the antennas directivity D, which is the ratio of the maximum power density in W/m2 to its average value over a sphere, as observed in the far field and considering a source with a Gaussian beam is

$$D = \frac{8}{\left[\frac{8}{3}-(cos\Delta\theta + cos^2\Delta\theta + \frac{1}{3}cos^3\Delta\theta)\right]} \quad (3)$$

where $\Delta\theta$ is the Guassian beam width of the nano-antenna.

*B.    Absorption Loss*

In the THz band, the biological molecules are excited by the EM waves at specific frequencies, that cause a vibration of these molecules [7]. Therefore, some of the propagating wave energy is transformed to kinetic energy and in other words to loss. Hence, absorption loss can be defined as the fraction of the incident EM waves which are not able to pass within the human tissue at a given frequency.

For an EM wave traveling a distance d and referring to Beer-Lambert law [8], the loss due to molecular absorption is given by

$$PL_{Abs} = -10log_{10}(e^{-\mu_{Abs}d}) \quad (4)$$

where $\mu_{Abs}$ is the molecular absorption coefficient which depends on the medium composition and it is given as

$$\mu_{Abs} = \frac{4\pi(n'')}{\lambda_g} \quad (5)$$

where $n''$ is the imaginary part of the tissue refractive index.

Since we are dealing with a huge number of water molecules for frequencies up to 1 THz, it is common to consider the complex permittivity using by the Debye Relaxation Model [9] [10] [11] as:

$$\epsilon = \epsilon' - j\epsilon'' \quad (6)$$

where $\epsilon'$ and $\epsilon''$ are the real and imaginary parts of the complex permittivity given by

$$\epsilon' = \epsilon_\infty + \frac{\epsilon_1-\epsilon_2}{1+(\omega\tau_1)^2} + \frac{\epsilon_2-\epsilon_\infty}{1+(\omega\tau_2)^2} \quad (7)$$

$$\epsilon'' = \frac{(\epsilon_1-\epsilon_2)(\omega\tau_1)}{1+(\omega\tau_1)^2} + \frac{(\epsilon_2-\epsilon_\infty)(\omega\tau_2)}{1+(\omega\tau_2)^2} \quad (8)$$

The values of $\epsilon_\infty, \epsilon_1, \epsilon_2, \tau_1$ and $\tau_2$ for the lower THz band are given in [9] and [11]. Consequently, $\epsilon'$ and $\epsilon''$ can be evaluated. On the other hand, $n''$ is given by

$$n'' = \sqrt{\frac{\sqrt{\epsilon'^2+\epsilon''^2}-\epsilon'}{2}} \quad (9)$$

Thus, $\mu_{Abs}$ and $PL_{Abs}$ can be calculated.

*C.    Scattering Loss*

The biological elements of the human body are of different types of composites, such as molecules and cells each having different geometry and different EM properties. Scattering by particles corresponds to the change of the direction of propagation and phase of the EM wave after interaction with the microscopic non-uniformities present in the human tissue. Scattering depends on the size, refractive index and shape of the individual particle as well as on the wavelength of the incident beam [12]. In this paper, we consider the scattering from small molecules as well as from large cells compared to the wavelength. Hence, the scattering path loss is

$$PL_{Sca} = -10log_{10}\left(e^{-(\mu_{small}+\mu_{large})d}\right) \quad (10)$$

where $\mu_{small}$ and $\mu_{large}$ are the scattering coefficients for small and large particles respectively and are defined as

$$\mu_{small} = \rho_v Q_{small}\sigma_g \text{ and } \mu_{large} = \rho_v Q_{large}\sigma_g \quad (11)$$

These coefficients depend on the particle concentration $\rho_v$ and the geometric cross section such that,

$$\rho_v = \frac{k}{\frac{4}{3}\pi r^3} \text{ and } \sigma_g = \pi r^2 \quad (12)$$

where $k$ is the volume fraction of the particle [13] and $r$ is the radius of the particle.

The absorption coefficient for small particles $Q_{small}$ depends on the scattering efficiency of small particle assuming a spherical absorbing particle modeled by Rayleigh scattering as

$$Q_{small} = \frac{8}{3}\psi^4 \text{Re}\left(\frac{n^2-1}{n^2+2}\right)^2 \quad (13)$$

where $\psi = 2\pi r/\lambda_g$ is the dimensionless size of the particle.

On the other hand, the absorption coefficient for large particles $Q_{large}$ depends on the scattering efficiency of large particles modeled using the anomalous diffraction approximation as

$$Q_{large} = 2 - \frac{4}{p}\sin(p) + \frac{4}{p^2}(1 - \cos(p)) - \frac{\sigma_{abs}}{\sigma_g} \quad (14)$$

where $p = 4\pi r(n-1)/\lambda$ is the phase delay of the wave passing through the center of the particle and $\sigma_{abs}$ is the molecular absorption cross section.

### III. SIMULATION RESULTS

In this section, we numerically evaluate the analytical models presented in Sec. II. However, we exclude the scattering loss since it is negligible compared to the absorption and spreading losses. Moreover, we take into account realistic parameters of the intrabody properties [9] [10] [11]. In this work, we conduct our analysis for water, skin and epidermis medium. In a fashion similar to [16-48], the work in this project is a continuation of the several previous contributions, which have been reported in the literature in various relevant technologies, systems, and their associated applications.

#### A. Path Loss and Frequency

Figs. 1 and 2 illustrate the total path loss in dB with respect to the medium wavelength for water, skin and epidermis respectively. Similar to conventional communication models in the megahertz frequency ranges, the path loss increases with frequency.

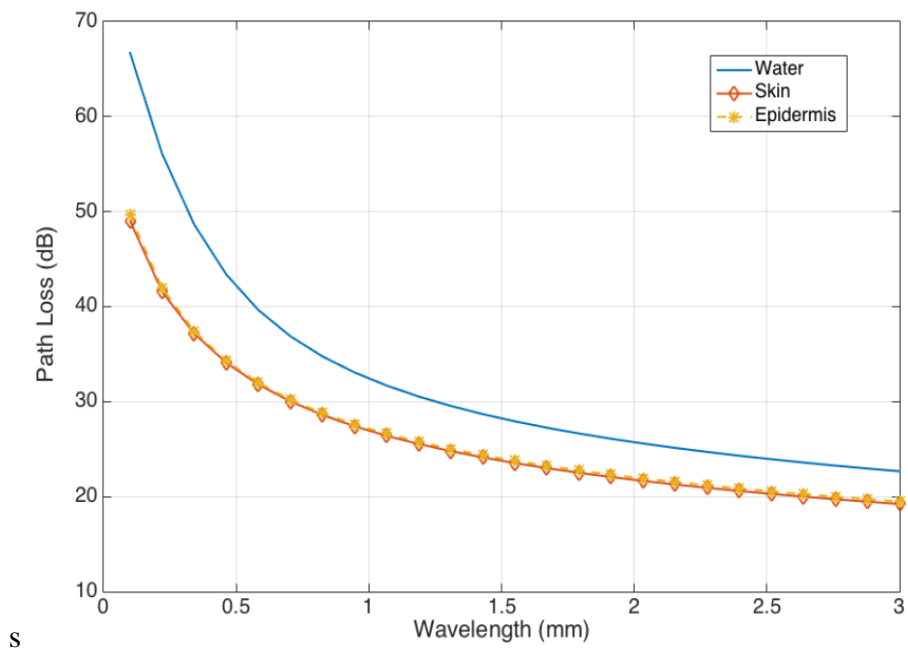

s

*Figure 1 Path loss (P$_L$), at d = 1mm and Δθ = 0.5 rad.*

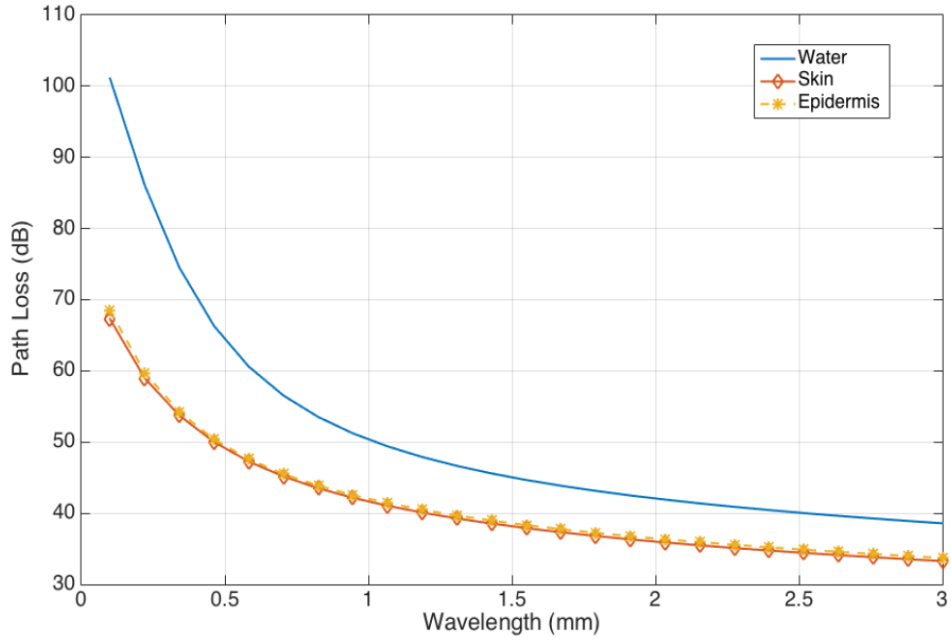

*Figure 2 Path loss ($P_L$), at d =2mm and Δθ = 0.5 rad.*

### B. Path Loss and Distance

Fig. 3 visualize the total path loss with respect to the communication distance d for water, skin and epidermis respectively. Similarly, the path loss increases with distance.

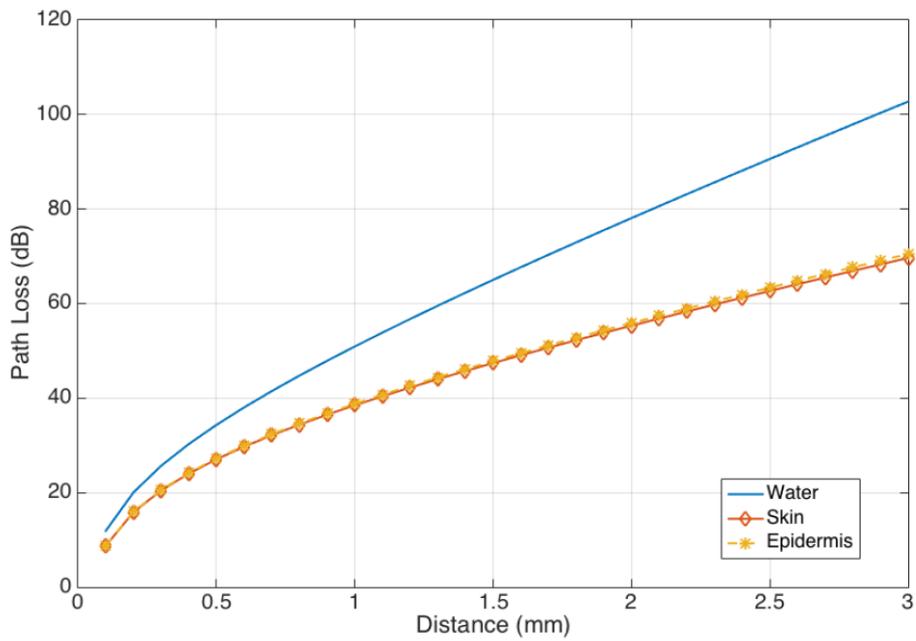

*Figure 3 Path loss ($P_L$), at $\lambda_0$ =0.3mm and Δθ = 0.5 rad.*

*C.    Path Loss and Nano-Antenna Directivity*

Fig. 4 shows the total path loss with respect to the Gaussian Beam Width $\Delta\theta$ for water, skin and epidermis respectively. From the graph, we can see that the path loss increases with large beam width and consequently, decreases with higher directivity.

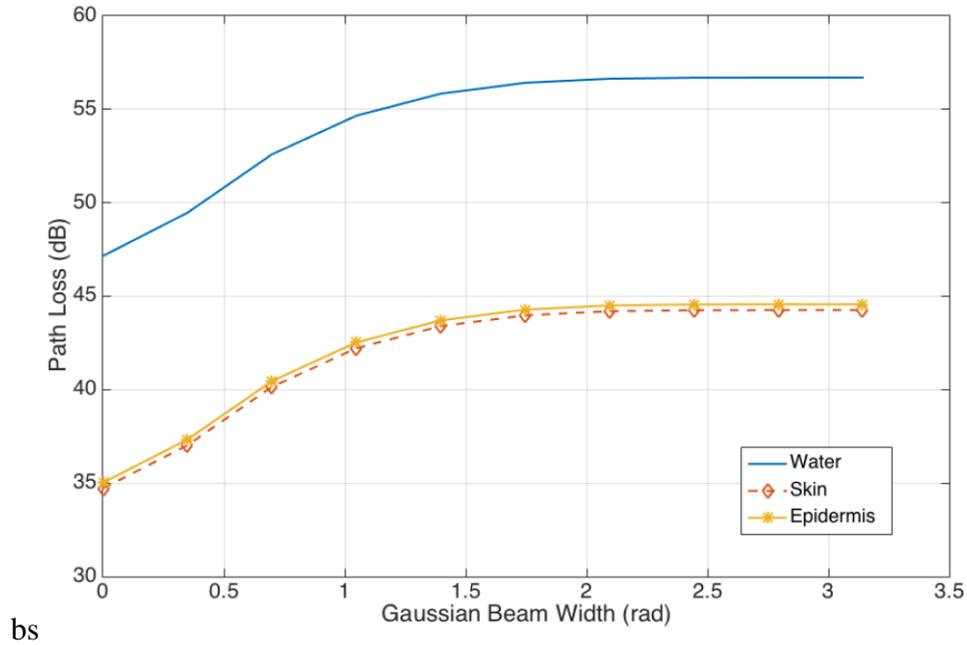

bs

*Figure 4 Path loss ($P_L$), at  $\lambda_0$=0.3mm and d = 1mm.*

## IV. DISCUSSION

From the numerical results shown in Sect. III, it is relevant that the path loss increases with frequency and distance.

The absorption loss is governed by the absorption coefficients of the human tissue molecules that is, referring to [9] and [14], are considerably high at THz frequencies. The reason behind this is the fact that the rotation transition of water molecules is located in this band.

The scattering loss depends on the radius of the human body particles. The size of the scatterers at THz being much smaller than the wavelength of the propagating THz wave [15], negligible scattering loss compared to the absorption loss will result. Therefore, the spreading and absorption losses are the main contributors to the total path loss at THz frequencies.

The relationship between path loss and communication distance is governed by the absorption loss which is, for water molecules, is larger than that for skin and epidermis at larger communication distance. That's why we can see approximately same path loss value at lower distances (<0.5 mm). However, when the communication distance increases, the curves diverge. The path loss in water (80 dB at 2 mm) is much higher than that of the skin and epidermis (55 dB at 2mm).

On the other hand, the nano-antenna directivity seems to have a big impact on the communication channel at THz frequencies. From the simulated values, we can see that the increasing the Gaussian beam width from 0 to 1.5 rad increases the path loss by 10 dB approximately for water, skin and epidermis. However, for a beam width between 1.5 and 3 rad, the path loss is approximately constant for each medium.

Therefore, we can see that the major impact is for the communication distance as well as the operating frequency in the human body. These two factors are more critical to consider than the nano-antenna directivity.

## V.   CONCLUSION

Nano-networks are exponentially expanding. The EM communication channel is having more attention and is promising to be adopted in WBANs. However, it needs deep study of its characteristics. In short, the main factors that affect the EM wave traveling in-vivo at THz frequencies are the nano-antenna directivity, the communication distance and the operating frequency. However, the major impact is that of distance and frequency.  In a fashion similar to [16]-[48], the work in this project is a continuation of the several previous contributions, which have been reported in the literature in various relevant technologies, systems, and their associated applications.

# REFERENCES

[1]    I.F. Akyildiz, "Nanonetworks: A new frontier in communications," *Proceedings of the 2010 International Conference on Security and Cryptography (SECRYPT),* July 2010, pp. IS–5–IS–5.

[2]    H. Guo, P. Johari, J. Jornet, and Z. Sun, "Intra-body optical channel modeling for in-vivo wireless nanosensor networks," 2015.

[3]    G. E. Santagati, T. Melodia, L. Galluccio, and S. Palazzo, "Ultrasonic networking for e-health applications," *IEEE Wireless Communications*, vol. 20, no. 4, pp. 74–81, 2013.

[4]    I.F. Akyildiz, F. Fekri, R. Sivakumar, C. R. Forest, and B. K. Hammer, "Monaco: fundamentals of molecular nano-communication networks," *IEEE Wireless Communications*, vol. 19, no. 5, pp. 12–18, 2012.

[5]    J.M. Jornet and I.F. Akyildiz, "Channel modeling and capacity analysis for electromagnetic wireless nanonetworks in the terahertz band," *IEEE Transactions on Wireless Communications*, vol. 10, no. 10, pp. 3211– 3221, 2011.

[6]    C. A. Balanis, *Antenna theory: analysis and design*. John Wiley & Sons, 2016.

[7]    P. Gans, *Vibrating molecules*. Chapman and Hall London, 1971.

[8]    K. Fuwa and B. Valle, "The physical basis of analytical atomic absorp- tion spectrometry. the pertinence of the beer-lambert law." *Analytical Chemistry*, vol. 35, no. 8, pp. 942–946, 1963.

[9]    K. Yaws, D. Mixon, and W. Roach, "Electromagnetic properties of tissue in the optical region," in *Biomedical Optics (BiOS) 2007*. International Society for Optics and Photonics, 2007, pp. 643 507–643 507.

[10]   J. Kindt and C. Schmuttenmaer, "Far-infrared dielectric properties of polar liquids probed by femtosecond terahertz pulse spectroscopy," *The Journal of Physical Chemistry*, vol. 100, no. 24, pp. 10 373–10 379, 1996.

[11]   J. Xu, K. W. Plaxco, and S. J. Allen, "Probing the collective vibra- tional dynamics of a protein in liquid water by terahertz absorption spectroscopy," *protein Science*, vol. 15, no. 5, pp. 1175–1181, 2006.

[12]   B. Chu, *Laser light scattering*. Elsevier, 1974.

[13]   Zohdi, T. I., F. A. Kuypers, and W. C. Lee. 2010. "Estimation of Red Blood Cell Volume Fraction from Overall Permittivity Measurements." International Journal of Engineering Science 48 (11): 1681–91.

[14]   C. B. Reid, G. Reese, A. P. Gibson, and V. P. Wallace, "Terahertz time- domain spectroscopy of human blood," *IEEE journal of biomedical and health informatics*, vol. 17, no. 4, pp. 774–778, 2013.

[15]   H. P. Erickson, "Size and shape of protein molecules at the nanometer level determined by sedimentation, gel filtration, and electron mi- croscopy," *Biological procedures online*, vol. 11, no. 1, p. 32, 2009

[16]   A. Omar and R. Shubair, "UWB coplanar waveguide-fed-coplanar strips spiral antenna," in 2016 10th European Conference on Antennas and Propagation (EuCAP), Apr. 2016, pp. 1–2.

[17]   H. Elayan, R. M. Shubair, J. M. Jornet, and P. Johari, "Terahertz channel model and link budget analysis for intrabody nanoscale communication," IEEE transactions on nanobioscience, vol. 16, no. 6, pp. 491–503, 2017.

[18]   H. Elayan, R. M. Shubair, and A. Kiourti, "Wireless sensors for medical applications: Current status and future challenges," in Antennas and Propagation (EUCAP), 2017 11th European Conference on. IEEE, 2017, pp. 2478–2482.

[19]  H. Elayan and R. M. Shubair, "On channel characterization in human body communication for medical monitoring systems," in Antenna Technology and Applied Electromagnetics (ANTEM), 2016 17th International Symposium on. IEEE, 2016, pp. 1–2.

[20]  H. Elayan, R. M. Shubair, A. Alomainy, and K. Yang, "In-vivo terahertz em channel characterization for nano-communications in wbans," in Antennas and Propagation (APSURSI), 2016 IEEE International Symposium on. IEEE, 2016, pp. 979–980. 35.

[21]  H. Elayan, R. M. Shubair, and J. M. Jornet, "Bio-electromagnetic thz propagation modeling for in-vivo wireless nanosensor networks," in Antennas and Propagation (EUCAP), 2017 11th European Conference on. IEEE, 2017, pp. 426–430.

[22]  H. Elayan, C. Stefanini, R. M. Shubair, and J. M. Jornet, "End-to-end noise model for intra-body terahertz nanoscale communication," IEEE Transactions on NanoBioscience, 2018.

[23]  H. Elayan, P. Johari, R. M. Shubair, and J. M. Jornet, "Photothermal modeling and analysis of intrabody terahertz nanoscale communication," IEEE transactions on nanobioscience, vol. 16, no. 8, pp. 755–763, 2017.

[24]  H. Elayan, R. M. Shubair, J. M. Jornet, and R. Mittra, "Multi-layer intrabody terahertz wave propagation model for nanobiosensing applications," Nano Communication Networks, vol. 14, pp. 9–15, 2017.

[25]  H. Elayan, R. M. Shubair, and N. Almoosa, "In vivo communication in wireless body area networks," in Information Innovation Technology in Smart Cities. Springer, 2018, pp. 273–287.

[26]  M. O. AlNabooda, R. M. Shubair, N. R. Rishani, and G. Aldabbagh, "Terahertz spectroscopy and imaging for the detection and identification of illicit drugs," in Sensors Networks Smart and Emerging Technologies (SENSET), 2017, 2017, pp. 1–4. 36.

[27]  M. S. Khan, A.-D. Capobianco, A. Iftikhar, R. M. Shubair, D. E. Anagnostou, and B. D. Braaten, "Ultra-compact dual-polarised UWB MIMO antenna with meandered feeding lines," IET Microwaves, Antennas & Propagation, vol. 11, no. 7, pp. 997–1002, Feb. 2017.

[28]  M. S. Khan, A. Capobianco, S. M. Asif, D. E. Anagnostou, R. M. Shubair, and B. D. Braaten, "A Compact CSRR-Enabled UWB Diversity Antenna," IEEE Antennas and Wireless Propagation Letters, vol. 16, pp. 808–812, 2017.

[29]  R. M. Shubair, A. M. AlShamsi, K. Khalaf, and A. Kiourti, "Novel miniature wearable microstrip antennas for ISM-band biomedical telemetry," in 2015 Loughborough Antennas Propagation Conference (LAPC), 2015, pp. 1–4.

[30]  R. M. Shubair, "Robust adaptive beamforming using LMS algorithm with SMI initialization," in 2005 IEEE Antennas and Propagation Society International Symposium, vol. 4A, Jul. 2005, pp. 2–5 vol. 4A.

[31]  F. A. Belhoul, R. M. Shubair, and M. E. Ai-Mualla, "Modelling and performance analysis of DOA estimation in adaptive signal processing arrays," in 10th IEEE International Conference on Electronics, Circuits and Systems, 2003. ICECS 2003. Proceedings of the 2003, vol. 1, Dec. 2003, pp. 340–343 Vol.1.

[32]  E. Al-Ardi, R. Shubair, and M. Al-Mualla, "Direction of arrival estimation in a multipath environment: an overview and a new contribution," in ACES, vol. 21, no. 3, 2006.

[33]  G. Nwalozie, V. Okorogu, S. Maduadichie, and A. Adenola, "A simple comparative evaluation of adaptive beam forming algorithms," International Journal of Engineering and Innovative Technology (IJEIT), vol. 2, no. 7, 2013.

[34]  M. Bakhar and D. P. Hunagund, "Eigen structure based direction of arrival estimation algorithms for smart antenna systems," IJCSNS International Journal of Computer Science and Network Security, vol. 9, no. 11, pp. 96–100, 2009.

[35]   M. Al-Nuaimi, R. Shubair, and K. Al-Midfa, "Direction of arrival estimation in wireless mobile communications using minimum variance distortionless response," in The Second International Conference on Innovations in Information Technology (IIT'05), 2005.

[36]   R. M. Shubair and W. Jassmi, "Performance analysis of optimum SMI beamformers for spatial interference rejection," in 2006 IEEE International Symposium on Circuits and Systems, May 2006, pp. 4 pp.–4746.

[37]   R. M. Shubair and H. Elayan, "In vivo wireless body communications: State-of-the-art and future directions," in Antennas & Propagation Conference (LAPC), 2015 Loughborough. IEEE, 2015, pp. 1–5.

[38]   R. M. Shubair and W. Jessmi, "Performance analysis of SMI adaptive beamforming arrays for smart antenna systems," in 2005 IEEE Antennas and Propagation Society International Symposium, vol. 1B, 2005, pp. 311–314 vol. 1B.

[39]   R. M. Shubair and A. Al-Merri, "Robust algorithms for direction finding and adaptive beamforming: performance and optimization," in The 2004 47th Midwest Symposium on Circuits and Systems, 2004. MWSCAS '04, vol. 2, Jul. 2004, pp. II–589–II–592 vol.2. 25.

[40]   R. M. Shubair and A. Merri, "A convergence study of adaptive beamforming algorithms used in smart antenna systems," in 11th International Symposium on Antenna Technology and Applied Electromagnetics [ANTEM 2005], Jun. 2005, pp. 1–5.

[41]   R. M. Shubair, "Improved smart antenna design using displaced sensor array configuration," Applied Computational Electromagnetics Society Journal, vol. 22, no. 1, p. 83, 2007. 27

[42]   R. M. Shubair and R. S. A. Nuaimi, "A Displaced Sensor Array Configuration for Estimating Angles of Arrival of Narrowband Sources under Grazing Incidence Conditions," in 2007 IEEE International Conference on Signal Processing and Communications, Nov. 2007, pp. 432–435.

[43]   A. Hakam and R. M. Shubair, "Accurate detection and estimation of radio signals using a 2d novel smart antenna array," in 2013 IEEE 20th International Conference on Electronics, Circuits, and Systems (ICECS), Dec. 2013, pp. 807–810.

[44]   R. M. Shubair and H. Elayan, "Enhanced WSN localization of moving nodes using a robust hybrid TDOA-PF approach," in 2015 11th International Conference on Innovations in Information Technology (IIT), Nov. 2015, pp. 122–127.

[45]   R. M. Shubair, K. AlMidfa, A. Al-Marri, and M. Al-Nuaimi, "Robust algorithms for doa estimation and adaptive beamforming in wireless mobile communications," International Journal of Business Data Communications and Networking (IJBDCN), vol. 2, no. 4, pp. 34–45, 2006.

[46]   F. O. Alayyan, R. M. Shubair, Y. H. Leung, A. M. Zoubir, and O. Alketbi, "On MMSE Methods for blind identification of OFDM based SIMO systems," in 2009 IFIP International Conference on Wireless and Optical Communications Networks, Apr. 2009, pp. 1–5.

[47]   H. Elayan and R. M. Shubair, "Towards an Intelligent Deployment of Wireless Sensor Networks," in Information Innovation Technology in Smart Cities. Springer, Singapore, 2018, pp. 235–250.

[48]   H. Elayan and R. M. Shubair, "Improved DV-hop localization using node repositioning and clustering," in 2015 International Conference on Communications, Signal Processing, and their Applications (ICCSPA), Feb. 2015, pp. 1–6.